# Experimental Test of the validity of "Isotropic" Approximation for the Mechanical Behaviour of Clay

## P. Evesque [1], M. Hattab [2]

[1] Lab MSSMat, UMR 8579 CNRS, Ecole Centrale Paris
92295 CHATENAY-MALABRY, France, e-mail evesque@mssmat.ecp.fr
[2] LPMM/ISGMP, UMR 7554 CNRS, Université de Metz, île de Saulcy,
57045 METZ cedex 01, France, hattab@lpmm.univ-metz.fr

**Abstract:**

*Experimental data from axially symmetric compression test at constant mean pressure $p=(\sigma_1+\sigma_2+\sigma_3)/3$ on kaolinite clay are used to study the validity of an "isotropic" modelling as a function of the overconsolidation ratio (OCR). The isotropic assumption is found to be quite good for $2<OCR<3$ and/or in the range of small deformation for $OCR>4$. For very large OCR (OCR >10), anisotropic response is observed at few percents axial deformation. Relation with anisotropic distribution of local force is made.*

**Pacs # :** 5.40 ; 45.70 ; 62.20 ; 83.70.Fn

___________________________________________________________________

Recent papers [1-3] have proposed a new simple modelling of the behaviour of granular media. It is based on an incremental modelling, assuming an "isotropic" response [4] governed by two plastic parameters, *i.e.* the pseudo Young modulus $1/C_o$ and pseudo Poisson coefficient $\nu$. This modelling assumes then that the incremental response of any compression test with axial symmetry obeys at any stage of the deformation the following equation:

$$\begin{pmatrix} d\varepsilon_1 \\ d\varepsilon_2 \\ d\varepsilon_3 \end{pmatrix} = -C_o \begin{pmatrix} 1 & -\nu & -\nu \\ -\nu & 1 & -\nu \\ -\nu & -\nu & 1 \end{pmatrix} \begin{pmatrix} d\sigma_1 \\ d\sigma_2 \\ d\sigma_3 \end{pmatrix} \quad (1)$$

Where $C_o$ plays the part of an inverse pseudo Young modulus and $\nu$ the part of a pseudo Poisson coefficient. $\sigma_i$ states for the effective stress supported by the granular assembly alone; it does not include the pressure $u_w$ supported directly by the liquid invading the pores. It has been proposed in [1-3] that evolution of $C_o$ and $\nu$ has to be determined from any test of triaxial compression with axial symmetry; and it has been found from triaxial compression at constant $\sigma_3=\sigma_2$ that $\nu$ depends only the stress ratio $\sigma_1/\sigma_3$.

   Indeed, this "isotropic" assumption may be a crude approximation even when the medium remains homogeneous with axial symmetry, since as the axial deformation proceeds, the distribution of contacts evolves which can induce anisotropy; in turn, this anisotropy modifies the response equation, *i.e.* Eq. (1).

For instance, still assuming that the granular medium remains homogeneous with axial symmetry during the whole deformation, one expects the more general form to be valid:

$$\begin{pmatrix} d e_1 \\ d e_2 \\ d e_3 \end{pmatrix} = -C_o \begin{pmatrix} 1 & -\nu' & -\nu' \\ -\nu & a & -\nu'' \\ -\nu & -\nu'' & a \end{pmatrix} \begin{pmatrix} d s_1 \\ d s_2 \\ d s_3 \end{pmatrix} \quad (2)$$

At this stage, it is worth recalling that $\nu=\nu'$ if *the* two different compression paths ($\delta\sigma_1 \neq 0, \delta\sigma_2=0, \delta\sigma_3=0$) and ($\delta\sigma_1=0, \delta\sigma_2 \neq 0, \delta\sigma_3=0$) pertain to the same incremental linear zone. It is also worth recalling that any axial compression give access to the quantity $\alpha-\nu''$ only so that $\alpha$ and $\nu''$ cannot be measured separately with such axi-symmetric triaxial apparatus.

Anyhow, one can then ask whether it is useful to use Equation (2) rather than its simplified "isotropic" version, *i.e.* Eq. (1). Indeed this can be checked directly from experimental data. For instance the validity of the "isotropic" version has been discussed when applied to oedometer compression [3,5]. But this validity can be questioned in a more general way. One way to check its validity is to apply an axially symmetric triaxial compression at constant pressure $p=(\sigma_1+\sigma_2+\sigma_3)/3$, since one expects $\delta v=0$ in the case of an "isotropic" response whatever the applied deviatoric stress $q=\sigma_1-\sigma_2$.

However to investigate the variation of the specific volume v versus the deviatoric stress q and shows that v depends very little on q is not sufficient since applying Eqs. (1) or (2) allows to determine the variation law which is expected from the two approaches and to show that both contain the products $\delta\varepsilon_v = C_o \cdot f(\nu,\nu',\nu'') \cdot \delta q$ which can be small when either $C_o$, or $f$, or both are small. So it does not demonstrate that f is small. This is why one can use the same procedure as proposed in [6].

One is then faced to determine directly the function f. A better way to proceeds is just to use experimental plots of variations of $\varepsilon_v$ as a function of $\varepsilon_1$ or $\varepsilon_d$, where. $\varepsilon_d = \varepsilon_1 - \varepsilon_2$. Since dp=0 imposes $\delta\sigma_1=(2/3)\,\delta q=-2\delta\sigma_2=-2\delta\sigma_3$, one gets from Eq. (2),:

$$\delta\varepsilon_v = \delta\varepsilon_d\, 2[(1-\alpha)+(\nu'+\nu''-2\nu)]/(2+\alpha+2\nu+2\nu'+2\nu'-\nu'') \quad (3)$$

When the response is isotropic, one gets $\nu=\nu'=\nu''$ and $\alpha=1$, so Eq. (3) becomes :

$$\delta\varepsilon_v = \delta\varepsilon_d\, 2[O(1-\alpha)+O(\nu'+\nu''-2\nu)]/(2+\alpha+2\nu+2\nu'-\nu'') \quad (4)$$

where O(x) is a function which tends to 0 as x. Furthermore, an estimate of $\alpha$, $\nu$, $\nu'$ & $\nu''$ are 1, ½, ½ & ½ respectively, since these values are those obtained for the critical state. So, $\alpha=1$ and Eq. (4) becomes

$$\delta\varepsilon_v = (4/9)[O(1-\alpha)+O(\nu'-\nu'')]\,\delta\varepsilon_d \quad (5)$$

So Eq. (5) gives the expected variations of $\delta\varepsilon_v$ *vs.* $\delta\varepsilon_d$ for an axially symmetric compression at constant mean pressure p. It indicates that the smaller the slope of $\varepsilon_v$ *vs.* $\varepsilon_d$ the better the "isotropic" approximation. We report on Fig. 1 variations of $\varepsilon_v$ *vs.* $\varepsilon_d$ under constant-pressure-p compression obtained on Kaolinite clay at different over-consolidation ratio (OCR), ranging from OCR=1 to OCR=50. These tests have been performed under the following procedure: First, each sample of clay has been compressed at $p_o=1$ MPa; then the mean pressure p has been reduced to $p=p_o/OCR$; which is the mean pressure p which has been used during the axial compression.

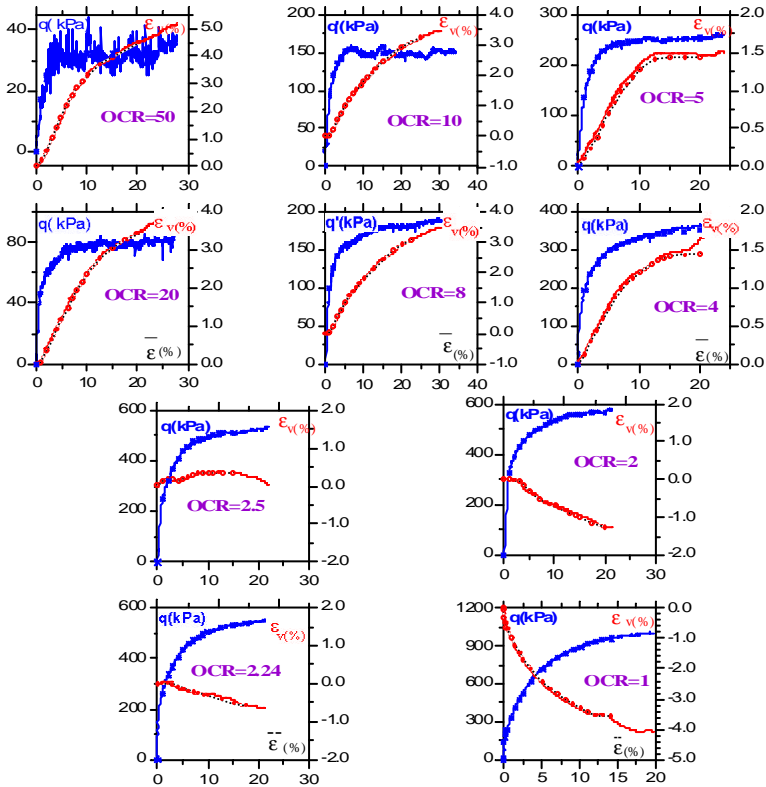

***Figure 1:*** *variations of the deviatoric stress* q *(crosses) and of volume deformation* $\varepsilon_v$ *(circles) vs. the deviatoric deformation* $\varepsilon_d$ *during an axial compression at constant mean pressure* p, *for different* p *value, but on the same Kaolinite clay. All samples were compacted initially at the same over-consolidation pressure* $p_o=1$ *MPa; each test corresponds to a different over-consolidation ratio* (OCR)*: OCR=$p_o$/p. OCR: 1, 2, 2.24, 2.5, 4, 5, 8, 10, 20, 50. Data from [7]* .

The $\varepsilon_v$ *vs.* $\varepsilon_d$ variations of Fig. 1 show that :
- an horizontal tangent at $\varepsilon_d=0$, except for OCR=$p_o$/p = 1, 4, 5 and perhaps at OCR= 50. This indicates that the response is rather "isotropic" at small $\varepsilon_d$ and small q/p hence, if one excepts OCR=$p_o$/p= 1, 4, 5 & 50.

- for OCR larger than 3 one observes an increase of the slope as soon as the deformation $\varepsilon_d$ overpasses 1%.

We report on Table 1 the maximum value of the slope $\varepsilon_v/\varepsilon_d$ for the different OCR. These data shows that the response is quite "isotropic" when OCR ranges in between 2 & 3. It is never isotropic for OCR=1. For other OCR "isotropic" approximation is good in the vicinity, *i.e.* few percent, of $\varepsilon_d=0$, except for OCR=4 & 5. We find also that the larger the OCR the more anisotropic the response since the larger the mean slope of $\varepsilon_v$ *vs.* $\varepsilon_d$. The slope $\delta\varepsilon_v/\delta\varepsilon_d$ can be as large as 0.37 which corresponds to a slope $\delta\varepsilon_v/\delta\varepsilon_1 \approx 0.45$.

It is worth mentioning that this last point is indeed not an obvious answer, because the volume variation seems to remain small in a large range of q/p when plotting of $\varepsilon_v$ *vs.* q or q/p. It is due to the fact that $C_o$ decreases strongly when increasing the OCR so that the amplitude of both $\varepsilon_d$ **and of** $\varepsilon_v$ deformations decreases strongly with q/p at large OCR. In other terms, it means simply that E, the pseudo Young modulus, increases strongly when OCR increases.

A consequence of this is that the procedure proposed in [6] to demonstrate the existence of a bifurcation process during the undrained test, which is based on the analysis of $\varepsilon_v$ *vs.* q curves obtained from compression tests at constant pressure p, does not assume or does not imply an isotropic response: this demonstration requires only that volume variation remains small in a range of q/p ratio laying around q/p=M', which is observed experimentally indeed.

| OCR=$p_o$/p | 1 | 2 | 2.24 | 2.5 | 4 | 5 | 8 | 10 | 20 | 50 |
|---|---|---|---|---|---|---|---|---|---|---|
| Max($\varepsilon_v/\varepsilon_d$) | -1 ? | -0.064 | 0.04 | 0.016 | 0.14 | 0.14 | 0.22 | 0.20 | 0.27 | 0.37 |
| slope at $\varepsilon_d=0$ | $-\infty$ ? | $\approx 0$ | $\approx 0$ | $\approx 0$ | ? | ? | $\approx 0$ | $\approx 0$ | $\approx 0$ | ? |

**Table 1:** maximum slope of $\varepsilon_v$ vs. $\varepsilon_d$ and slope of $\varepsilon_v$ vs. $\varepsilon_d$ at the origin of deformation for the different axially symmetric compression test at constant pressure p, for the same kaolinite clay for the same over-consolidation pressure $p_o$=1 M Pa, but at different over-consolidation ratio (OCR=$p_o$/p); *data from* [7].

This demonstrates that generation of anisotropy does not require the evolution of contacts, in clays at least, since anisotropic behaviours appear at small $\varepsilon_d$ already. This may be the experimental proof of the existence of a process able to generate a pure stress-induced anisotropy, without the requirement of strain generation. This result may then reinforce the results from numerical simulations on granular media obtained by Radjai and coworkers [8] who have observed two force networks, one which remains isotropic and corresponds to an isotropic answer, and the other which is the answer of the granular medium to an anisotropic stress [8] and which generates an anisotropic response. This is important to remark since some previous papers of one of us [2,3,5] was assuming that development of anisotropic answer was requiring the generation of an important axial strain ($\varepsilon_1$>5%). On the contrary, the experimental data on clays seems to infirm this hypothesis.

At last, it is worth noting a final point, which concerns the q *vs.* $\varepsilon_d$ curves: Fig. 1 data demonstrates that q tends to a limit value such that q/p≈1.5 – 1.6 at large deformation and for large OCR (OCR>5). Furthermore, the larger the OCR the sooner, *i.e.* the smaller $\varepsilon_d$ value at which, this q/p=1.5 limit value is reached. No doubt, these facts shall be attributed to the divergence of $C_o$ , (or E→0). So considering Eq. (2) and combining it with the experimental condition $\delta p=0$ impose that $\delta\varepsilon_d=[C_o/3]\delta q(2+\alpha+2\nu+2\nu'-\nu'')$ and $\delta\varepsilon_v=[C_o/3]\delta q(2-2\alpha-4\nu+2\nu'+2\nu'')$. In other words, the increase of Co imposes the increase of the volume variation; this is why one observes a large volume variation at large q/p in these large-OCR cases. It is also worth noting that this q/p ratio is larger than the expected one from natural friction of clay, since M'=1.5 or 1.6 leads to 37°< $\varphi$ <39° , according to $\sin\varphi=3M'/(6+M')$. This typical value of M' which is larger than the typical one of the critical state is probably induced by the dilatancy mechanism which is still visible from the experimental data, so that one should expect a decrease of the q/p ratio at deformation $\varepsilon_d$ larger than 25%.

As a conclusion this paper shows from experimental data on axially symmetric compression of Kaolinite clay at constant pressure that "isotropic" response is valid (i) in the whole domain of deformation for 2<OCR<4 , (ii) in a small range of axial deformation near $\varepsilon_d=0$ for OCR >4.

*Acknowledgements:* We want to thank Prof. J. Biarez for helpful discussion and PE wants to thank CNES for partial funding.

## References

[1] P. Evesque, "A Simple Incremental Modelling of Granular-Media Mechanics" , *poudres & grains* **9**, 1-12, (1999)
[2] P. Evesque, "On undrained test using Rowe's relation and incremental modelling : Generalisation of the notion of characteristic state" , *poudres & grains* **8**, 1-11, (1999)
[3] P. Evesque, "Éléments de mécanique quasi statique des milieux granulaires mouillés ou secs" , *poudres & grains* **NS1**, 1-155, (2000)
[4] When "isotropic" is written in bracket, it means that the matrix which represents the incremental stress-strain law is isotropic. It does not mean that the real material behaviour is isotropic, since the material complete response is non linear: Assuming that the material behaves isotropically would mean that its total response function is determined by the deviatoric stress q and the mean pressure p' though a function f(q,p') which is non linear. From derivation of f against q and p', one finds the total deformation tensor and second derivation leads to the determination of the incremental response.
[5] P. Evesque, "On Jaky constant of oedometers, Rowe's relation and incremental modelling"*, poudres & grains* **6**, 1-9, (1999).
[6] P. Evesque & M. Hattab, "Experimental proof of the bifurcation process during the undrained test in granular materials", *poudres & grains* **12 (1)**, 1-4, (2000)
[7] M. Hattab, *Etude expérimentale du comportement dilatant des argiles surconsolidées*, thèse de l'Ecole centrale Paris (1995-38)
[8] F. Radjai, D. Wolf, M. Jean & J.J. Moreau, "Bimodal character of stress transmission in granular

packing", *Phys. Rev. Lett.* **90**,61, (1998) ; F. Radjai, D. Wolf, S. Roux, M. Jean & J.J. Moreau, "Force network in dense granular media" , in *Powders & Grains 1997,* R.P. Behringer & J.T. Jenkins eds, (Balkema, Rotterdam, 1997), pp. 211-214.

The electronic arXiv.org version of this paper has been settled during a stay at the Kavli Institute of Theoretical Physics of the University of California at Santa Barbara (KITP-UCSB), in june 2005, supported in part by the National Science Fundation under Grant n° PHY99-07949.

*Poudres & Grains* can be found at :
http://www.mssmat.ecp.fr/rubrique.php3?id_rubrique=402